\documentclass[prl,english,reprint,showpacs]{revtex4-1}

\usepackage{amsmath,amssymb,amsbsy,graphicx}
\usepackage{hyperref}

\usepackage{babel,natbib,mathrsfs}

\hypersetup{
    unicode=false,          
    pdftoolbar=true,        
    pdfmenubar=true,        
    pdffitwindow=false,     
    pdfstartview={FitH},    
    pdftitle={My title},    
    pdfauthor={Author},     
    pdfsubject={Subject},   
    pdfcreator={Creator},   
    pdfproducer={Producer}, 
    pdfkeywords={keyword1} {key2} {key3}, 
    pdfnewwindow=true,      
    colorlinks=true,       
    linkcolor=blue,          
    citecolor=blue,        
    filecolor=magenta,      
    urlcolor=blue           
}

\begin{document}

\title{Globally aligned states and hydrodynamic traffic jams in confined active suspensions}

\author{Adrien Lefauve} 
\author{David Saintillan} \email{dstn@illinois.edu.}
\affiliation{Department of Mechanical Science and Engineering, University of Illinois at Urbana-Champaign, Urbana, Illinois 61801, USA}

\date{\today}

\pacs{87.18.Hf, 47.63.Gd, 47.57.E-, 05.65.+b}

\begin{abstract}

Strongly confined active liquids are subject to unique hydrodynamic interactions due to momentum screening and lubricated friction by the confining walls. Using numerical simulations, we demonstrate that 2D dilute suspensions of fore-aft asymmetric polar swimmers in a Hele-Shaw geometry can exhibit a rich variety of novel phase behaviors depending on particle shape, including: coherent polarized density waves with global alignment, stationary asters, persistent counter-rotating vortices, density shocks and rarefaction waves.\ We also explain these phenomena using a linear stability analysis and a nonlinear traffic flow model, both derived from a mean-field kinetic theory.

\end{abstract}

\maketitle

The spontaneous emergence of collective motion, large-scale coherent patterns and complex unsteady flows is a generic feature of a wide variety of soft active systems \cite{Ramaswamy10,Saintillan13,Marchetti13}, including: bacterial swarms and suspensions \cite{Chen12,Cisneros11,Dunkel13}, reactive colloidal suspensions \cite{Ebbens10,Takagi13}, cytoplasmic extracts \cite{Woodhouse12}, solutions of biological filaments and motor proteins \cite{Schaller10,Sanchez12}, and shaken granular materials \cite{Kudrolli08,Deseigne10}. In these systems, an immersed microstructure (such as self-propelled particles or biofilament/motor-protein assemblies) converts a source of energy (often chemical) into microstructural changes that create active stresses and can drive coherent flows on mesoscopic length scales. Analyzing how the interactions between individual agents translate into collective dynamics in these dissipative nonequilibrium systems has been the subject of intense research in the past few years, and could pave the way for a deeper understanding of intracellular processes \cite{Woodhouse13} and enable new paradigms for the design of smart materials \cite{Hatwalne04,Voituriez05,Giomi11} and autonomous micromachines \cite{Darnton04,Sokolov10,DiLeonardo10}.

The prototypical case of a collection of motile particles in a viscous liquid has received much scrutiny. Experiments on microbial suspensions in the bulk  show that above a critical concentration spatiotemporally chaotic flows reminiscent of high-Reynolds-number turbulence spontaneously arise \cite{Dunkel13}, and are accompanied by a transition to local orientational order \cite{Cisneros11}. As demonstrated by particle simulations \cite{Hernandez05,Saintillan07} and kinetic theories \cite{Saintillan08,Subra09,Baskaran09}, this transition can be explained as a result of long-ranged hydrodynamic interactions (HI) between swimmers, which are driven by the force dipoles exerted by the particles on the medium owing to self-propulsion. These interactions drive a generic long-wave instability for the nematic order parameter in the form of bend modes in suspensions of so-called pushers or rear-actuated swimmers \cite{Baskaran09}. Recent evidence  suggests that direct steric interactions also  play a role in  dense systems \cite{Cisneros11,Dunkel13,Ezhilan13}.

The nature of HI between self-propelled particles, however, changes drastically under confinement. 
As a swimmer moves in a rigidly confined space, say in the thin gap $h$ between two parallel flat plates, the disturbance flow induced by its permanent force dipole is partially screened by the walls and decays rapidly as $\sim 1/r^{3}$ in the far field \cite{Liron76,Bhatta06,Brotto13}, where $r$ is the distance from the particle, as opposed to the usual $ 1/r^{2}$ decay in the absence of confinement.\ Under these conditions, the far-field flow assumes the Hele-Shaw form and can be modeled as a two-dimensional potential flow \cite{Bhatta06}: $\mathbf{u}(\mathbf{r})=-(h^{2}/12\eta)\nabla\Pi(\mathbf{r})$, where $\mathbf{r}=(x,y)$, $\eta$ is the dynamic viscosity, and $\{\mathbf{u},\Pi\}$ denote the gap-averaged velocity and pressure fields. As a particle with thickness $\lesssim h$ located at position $\mathbf{R}(t)$ moves in an external flow $\mathbf{u}(\mathbf{r},t)$ in such a geometry, the disturbance velocity it induces is now dominated by a mass dipole singularity with $1/r^{2}$ spatial decay resulting from the displacement of the fluid by the finite volume of the particle \cite{Brotto13,Beatus09}:
\vspace{-0.1cm}
\begin{equation}
\mathbf{u}^{d}(\mathbf{r}|\mathbf{R}(t),\boldsymbol{\sigma})=\frac{1}{2\pi|\mathbf{r}|^{2}}(2\hat{\mathbf{r}}\hat{\mathbf{r}}-\mathbf{I})\cdot\boldsymbol{\sigma}, \label{eq:dipole}\vspace{-0.1cm}
\end{equation}
where $\mathbf{r}$ points from the particle center-of-mass  $\mathbf{R}(t)$, and $\hat{\mathbf{r}}=\mathbf{r}/r$. The dipole strength is proportional to the relative velocity between the particle and background flow: $\boldsymbol{\sigma}=\sigma[\dot{\mathbf{R}}(t)-\mathbf{u}(\mathbf{R}(t))]$, where the prefactor $\sigma$ scales as the particle surface area in the $(x,y)$ plane.\  Another defining feature of strongly confined systems is the way particles respond to a prescribed flow field.\ Brotto \textit{et al.}\ \cite{Brotto13} recently explained that an anisotropic swimmer lacking fore-aft symmetry and undergoing lubricated friction with the walls will generally be subject to an anisotropic mobility and to alignment in both the flow and flow-gradient directions. Using a simple dumbbell model, they showed that the motion of a swimmer with position $\mathbf{R}$ and director $\mathbf{p}$ ($|\mathbf{p}|=1$) in a flow field $\mathbf{u}$ obeys
\begin{eqnarray} 
&&\dot{\mathbf{R}} = v_s \, \mathbf{p} + \mu_{\perp} \, (\mathbf{I}-\mathbf{pp}) \cdot  \mathbf{u} + \mu_{\parallel} \, \mathbf{pp} \cdot \mathbf{u}\label{motion_r}, \\
&&\dot{\mathbf{p}} = \nu \,  (\mathbf{I}-\mathbf{pp}) \cdot \mathbf{u}  + \nu' \, (\mathbf{I}-\mathbf{pp})\cdot \nabla \mathbf{u} \cdot \mathbf{p} \label{motion_p}.
\end{eqnarray}
Here, $v_{s}$ is the swimming speed, and $\mu_{\perp,\parallel}$ are the transverse and longitudinal  mobilities. While reorientation by the flow gradient (with constant $\nu'$) corresponds to the well-known Jeffery's equation, reorientation by the velocity itself is a consequence of confinement and  fore-aft asymmetry, with `large-head' swimmers ($\nu<0$)  aligning against the flow and `large-tail' swimmers ($\nu>0$)  with the flow \cite{Brotto13}. Based on these effects, Brotto \textit{et al.}\ derived a kinetic model for a 2D suspension of  interacting confined swimmers and predicted a novel long-wave linear instability of the homogeneous isotropic phase, amplifying perturbations in longitudinal polarization. However, the phase behavior and pattern formation that these instabilities lead to in the nonlinear regime in finite-sized systems remain unknown.

This Letter presents the first detailed description of these nonlinear dynamics via discrete particle simulations of rigidly confined 2D suspensions.\ We show that the distinctive features of HI and swimmer orientational dynamics under confinement lead to a rich variety of phase behaviors, including: coherent polarized density waves, metastable stationary asters, persistent counter-rotating vortices, density shocks and rarefaction waves.\ These phenomena, which are unobserved in unconfined bulk systems, are also substantiated by a finite-wavelength linear stability analysis and a nonlinear traffic flow model, both derived from a mean-field kinetic theory. 

We simulate the 2D dynamics of a large number $N$ of confined particles by integrating (\ref{motion_r})-(\ref{motion_p}) in time, where the velocity $\mathbf{u}$ is the superposition of the dipolar fields driven by the swimmers according to (\ref{eq:dipole}).\ We also impose rotational diffusion with diffusivity $d$ by adding to $\dot{\mathbf{p}}$ a vector $\sqrt{2d/\delta t}\,(\mathbf{I}-\mathbf{pp})\cdot\mathbf{n}$, where $\delta t$ is the time step and the components of $\mathbf{n}$ follow a normal distribution with zero mean and unit variance. Center-of-mass diffusion is not imposed, though the coupling between rotational diffusion and swimming naturally introduces spatial diffusion, with long-time diffusivity $D=v^{2}_{s}/2d$ in the isotropic phase \cite{Berg93}. The computational domain is a 2D square box of linear size $L$, and particle positions and orientations are initialized randomly following a uniform law. Dipolar interactions between particles and their periodic images are computed accurately using an efficient algorithm \cite{Lefauve13}.\ We scale time by the orientational relaxation timescale $d^{-1}$ and lengths by the distance $\ell=v_{s}/2d$ traveled by a swimmer before it loses memory of its orientation.\  We set $\mu_{\perp,\parallel}=1$, $\nu'=0$, and focus primarily on the influence of the polar alignment parameter $\nu$ and system size $L/\ell$,  as well as  area fraction $\phi=\sigma c_{0}/2$ (with $c_{0}=N/L^{2}$ the number density).\ Following Brotto \textit{et al.}\ and anticipating our results, we introduce a signed P\'eclet number $\mathrm{Pe}=2\nu\phi \ell$ comparing the rotation rate of a swimmer by the dipolar flows induced in the suspension and by rotational diffusion, and recall that long-wave polarization fluctuations are predicted to be unstable for large-head swimmers when $\mathrm{Pe}< -1$ \cite{Brotto13}. 

\begin{figure}[b]
  \vspace{-0.5cm}\centerline{\vspace{-0.3cm}
  \includegraphics[scale=0.5]{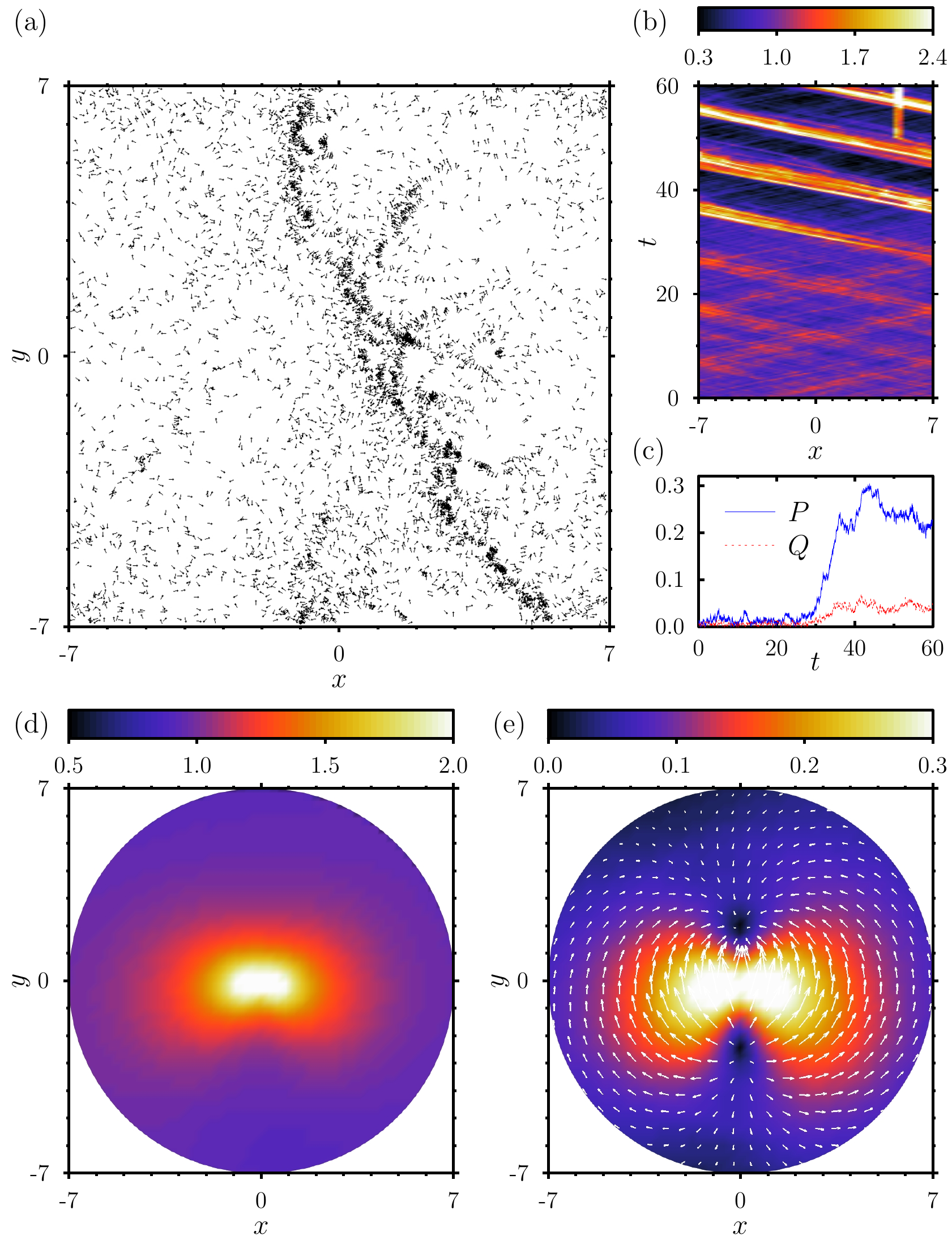}}
  \caption{(color online). Direct simulation with $N=5000$, $L/\ell=14$, $\phi=1\%$, and $\mathrm{Pe}=-2.2$. (a) Snapshot of the polarized density waves. (b) Spatiotemporal diagram of longitudinal concentration. (c) Global polar and nematic order parameters. (d) Pair distribution function $g(\mathbf{r})$. (e) Pair polarization $\mathbf{P}(\mathbf{r})$, with colors indicating $|\mathbf{P}|$. \label{fig:largehead}}
\end{figure}

A typical simulation in this regime is shown in Fig.~\ref{fig:largehead} and online video \cite{epaps} for $N=5000$, $\phi=1\%$, and $\mathrm{Pe}=-2.2$.\ Starting from uniform isotropy, Fig.\ \ref{fig:largehead}\hyperref[fig:largehead]{(a)} shows heavily polarized sharp density waves, which travel in arbitrary directions and exhibit a distinct curvature indicative of splay. The formation and growth of these waves is clearly seen on a spatiotemporal diagram of the longitudinal concentration  in Fig.~\ref{fig:largehead}\hyperref[fig:largehead]{(b)}, where swimmers are found to concentrate on a sharp front propagating at a characteristic speed $\approx 0.7 v_{s}$.\ In some instances, groups of particles also organize into stationary aster-shaped clusters, as evidenced by the vertical stripe forming at $t\approx 50$.\ To quantify particle alignment, we define a global polar order parameter $P=|\langle\mathbf{p}\rangle|$ (where $\langle\cdot\rangle$ is the suspension average) and a global nematic order parameter $Q$ as the positive eigenvalue of $\langle\mathbf{pp}-\mathbf{I}/2\rangle$. Both are shown in Fig.~\ref{fig:largehead}\hyperref[fig:largehead]{(c)} to grow from zero as the instability develops, and to plateau at long times when polarization appears to dominate.\ The emergence of global alignment is particularly surprising given the non-aligning nature of dipolar interactions and, to our knowledge, has never been observed before.
The structure of the density waves is further characterized in Fig.~\ref{fig:largehead}\hyperref[fig:largehead]{(d)}, showing the 2D pair distribution function $g(\mathbf{r})$, or probability of finding a second swimmer at position $\mathbf{r}$ if a first swimmer is located at the origin and points in the $+y$ direction.\ The anisotropy and curvature of the peak at the origin are consistent with the presence of longitudinal curved density waves.\ The pair polarization $\mathbf{P}(\mathbf{r})$ is also shown in Fig.~\ref{fig:largehead}\hyperref[fig:largehead]{(e)} and has the symmetry of a potential dipole flow.\ The direction of $\mathbf{P}(\mathbf{r})$, however, is opposite that of the flow driven by the particle at the origin, as one would expect for large-head swimmers that align against the flow ($\nu<0$).

\begin{figure}[b]
\centerline{\vspace{-0.4cm}
  \includegraphics[scale=0.5]{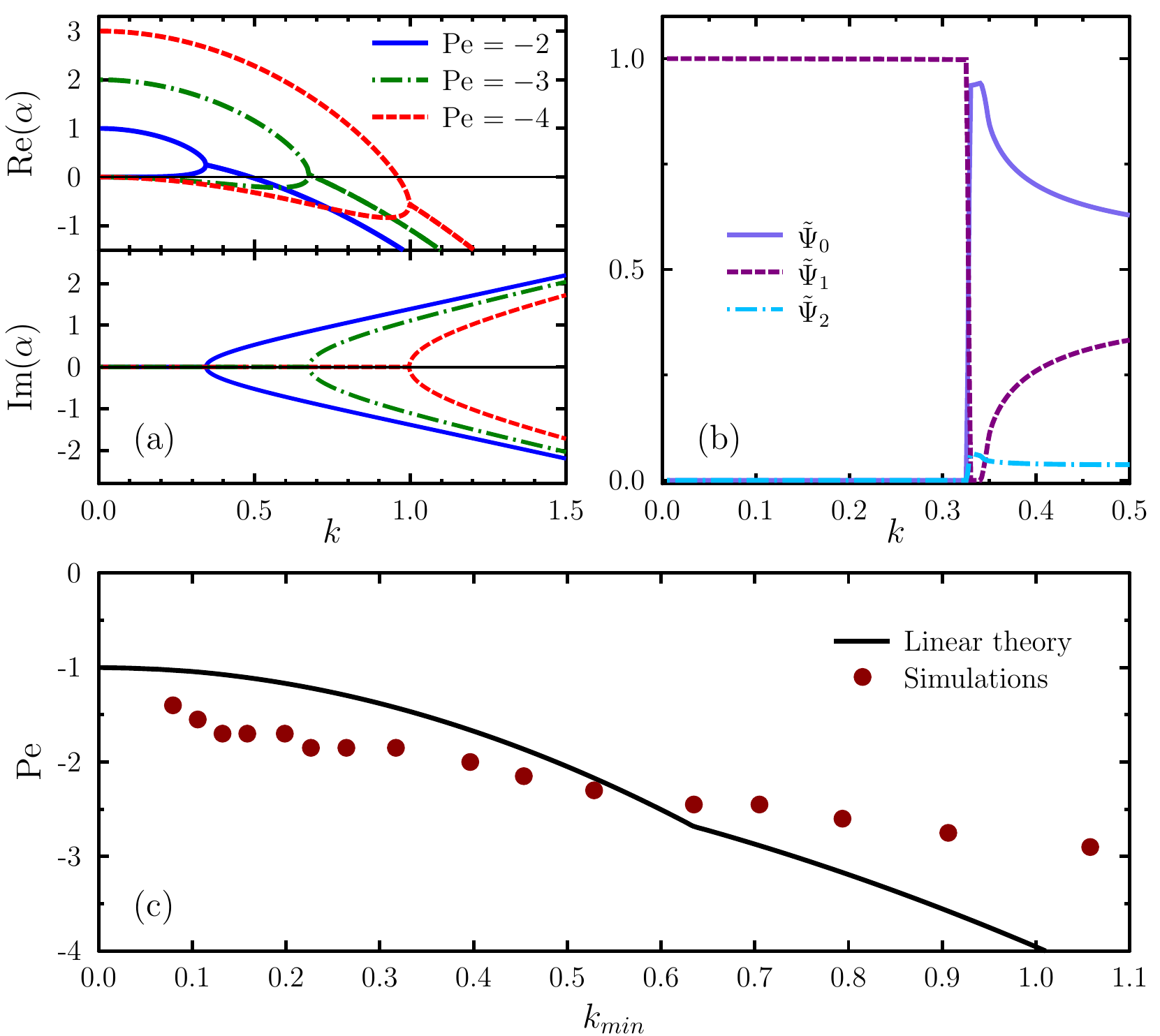}}
  \caption{(color online). (a) Real and imaginary parts of the growth rate $\alpha$ vs $k$ for various unstable values of $\mathrm{Pe}$. (b) Harmonic content of the unstable modes: zeroth, first and second harmonics vs $k$ for $\mathrm{Pe}=-2$. (c) Marginal P\'eclet number for instability vs wavenumber, from theory and  simulations.  \label{fig:stability}}
\end{figure}

To substantiate the observed pattern formation, we revisit the linear stability analysis of Brotto \textit{et al.}\ \cite{Brotto13} at finite wavenumber.\ The configuration of the suspension is described by the probability distribution $\Psi(\mathbf{r},\mathbf{p},t)$ of finding a particle at position $\mathbf{r}$ with orientation $\mathbf{p}$ at time $t$, which satisfies the Smoluchowski equation \cite{Saintillan08}
\begin{equation}
\partial_{t}\Psi+\nabla_{r}\cdot(\dot{\mathbf{R}}\Psi)+\nabla_{p}\cdot(\dot{\mathbf{p}}\Psi)=D\nabla_{r}^{2}\Psi+d\nabla_{p}^{2}\Psi, \label{smoluchowski}
\end{equation}
with $\dot{\mathbf{R}}$ and $\dot{\mathbf{p}}$ defined in (\ref{motion_r})-(\ref{motion_p}), where the fluid velocity  is now obtained as $\mathbf{u}(\mathbf{r},t)=\iint\Psi(\mathbf{r}',\mathbf{p},t)\mathbf{u}^{d}(\mathbf{r}|\mathbf{r}',\boldsymbol{\sigma})\,d\mathbf{r}'d\mathbf{p}$. We analyze the linear stability of the uniform isotropic state to a weak plane-wave perturbation with arbitrary wave vector $\mathbf{k}$: $\Psi(\mathbf{r},\mathbf{p},t)=c_{0}/2\pi+\varepsilon \tilde{\Psi}(\mathbf{p})\exp (i\mathbf{k}\cdot\mathbf{r}+\alpha t)$, where $\alpha$ is the complex growth rate and $|\varepsilon|\ll 1$.\ Linearization of (\ref{smoluchowski}) about this state yields an eigenvalue problem for $\{\alpha,\tilde{\Psi}\}$ that can be solved numerically \cite{Lefauve13}.\ The dispersion relation $\alpha(k)$ is shown in Fig.\ \ref{fig:stability}\hyperref[fig:stability]{(a)}, where a positive growth rate is seen to occur when $\mathrm{Pe}<-1$ for sufficiently small wavenumbers.\ In the limit of $k\rightarrow 0$, we recover the long-wave result of Brotto \textit{et al.}\ \cite{Brotto13}: $\mathrm{Re}(\alpha)=-(1+\mathrm{Pe})$.\ The unstable eigenmodes are illustrated in Fig.\ \ref{fig:stability}\hyperref[fig:stability]{(b)}, showing the zeroth, first and second Fourier coefficients of $\tilde{\Psi}(\theta)$ where $\theta=\mathrm{cos}^{-1} (\mathbf{p}\cdot\hat{\mathbf{k}})$, corresponding to fluctuations in concentration, polar and nematic splay alignments, respectively.\ At very low $k$, the instability only pertains to polarization in agreement with Brotto \textit{et al.}, but if $-3<\mathrm{Pe}<-1$ there also exists a finite range of wavenumbers for which the unstable eigenmodes couple all three harmonics, and propagate spatially as shown by the non-zero value of $\mathrm{Im}(\alpha)$.\ This is consistent with the observations on Fig.\ \ref{fig:largehead} ($\mathrm{Pe}=-2.2$, $k\ge 0.44$), and the ratio of $\tilde{\Psi}_{1}$ and $\tilde{\Psi}_{2}$ in Fig.\,\ref{fig:stability}\hyperref[fig:stability]{(b)} also mirrors that of $P$ and $Q$ in the nonlinear regime.\ A more quantitative comparison with the simulations is provided in Fig.~\ref{fig:stability}\hyperref[fig:stability]{(c)}, showing the marginal P\'eclet number for instability vs $k_{\mathrm{min}}=\sqrt{2}\pi\ell/L$, or  smallest wavenumber in a system of size $L$.\ Good agreement is found, with some level of discrepancy which we attribute to  the discrete nature of the simulations, finite amplitude of the fluctuations in the initial random condition, and nonlinearities. 

\begin{figure}[t]
\centerline{\vspace{-0.3cm}
  \includegraphics[scale=0.51]{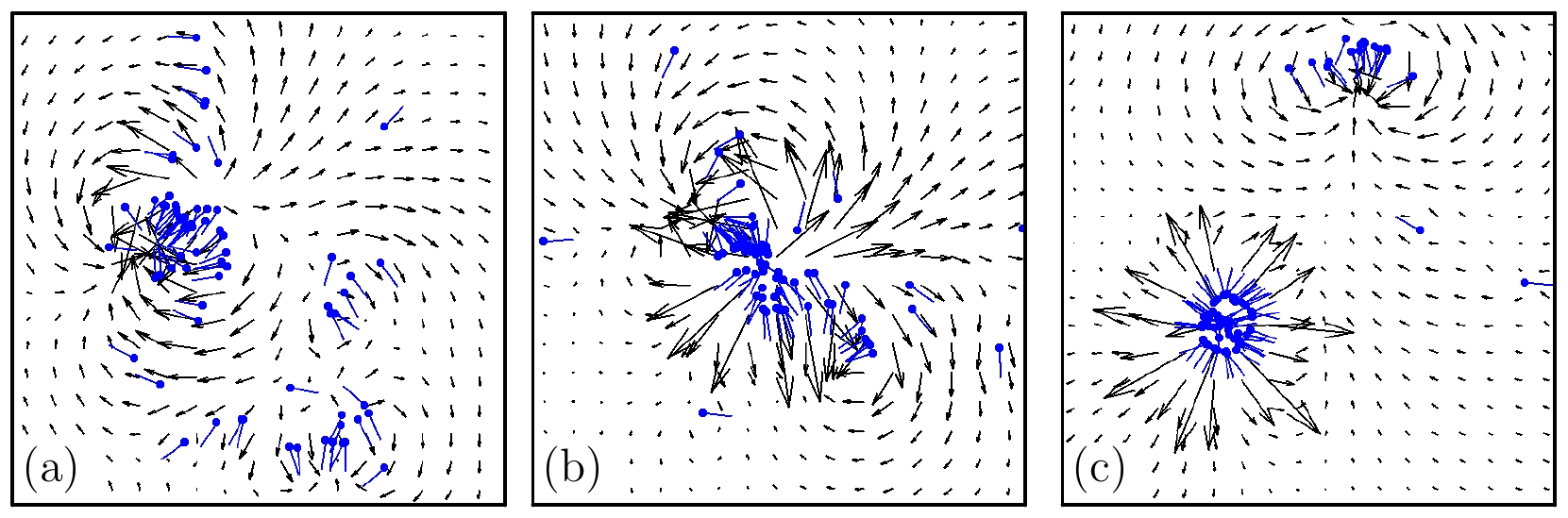}}
  \caption{(color online). Time sequence of particle positions and velocity field during the formation of a circular aster.\label{fig:aster}}\vspace{-0.4cm}
\end{figure}

The formation of stationary aster-shaped clusters mentioned above, which is not predicted by the linear analysis, is illustrated more clearly in Fig.\ \ref{fig:aster} and online video \cite{epaps}.\ As a finite-size polarized wave packet forms and propagates, it drives a flow akin to a large-scale potential dipole; see Fig.\ \ref{fig:aster}\hyperref[fig:aster]{(a)}.\ The propensity of large-head swimmers to align and swim against the flow causes the reorientation and capture of particles surrounding the wave as in Fig.\ \ref{fig:aster}\hyperref[fig:aster]{(b)}, and often leads to the formation of circular asters composed of concentric layers of converging swimmers; see Fig.\ \ref{fig:aster}\hyperref[fig:aster]{(c)}.\ As the diverging radial flow driven by the particles overcomes their swimming speed and captures neighboring swimmers, these clusters stabilize and continue to grow, sometimes resulting in long-lived multilayered structures.\ An analytical solution for $M$ radially-aligned swimmers at the vertices of a regular polygon shows that the equilibrium radius of such clusters depends linearly on particle size and number: $R=\sqrt{\sigma(M^{2}-1)/24\pi}\approx 0.115 \sqrt{\sigma}M$.\ Numerical experiments show that the clusters are weakly unstable, though they destabilize and break up on very long time scales. The formation of such asters has also been predicted in other active systems \cite{Kruse04,Gopinath12,Farrell12}, and notably in suspensions of swimmers with density-dependent motility \cite{Farrell12}, where the decrease in motility with concentration plays a similar role as the velocity slowdown induced here by HI.

\begin{figure}[t]
 \centerline{\vspace{-0.3cm}
  \includegraphics[scale=0.5]{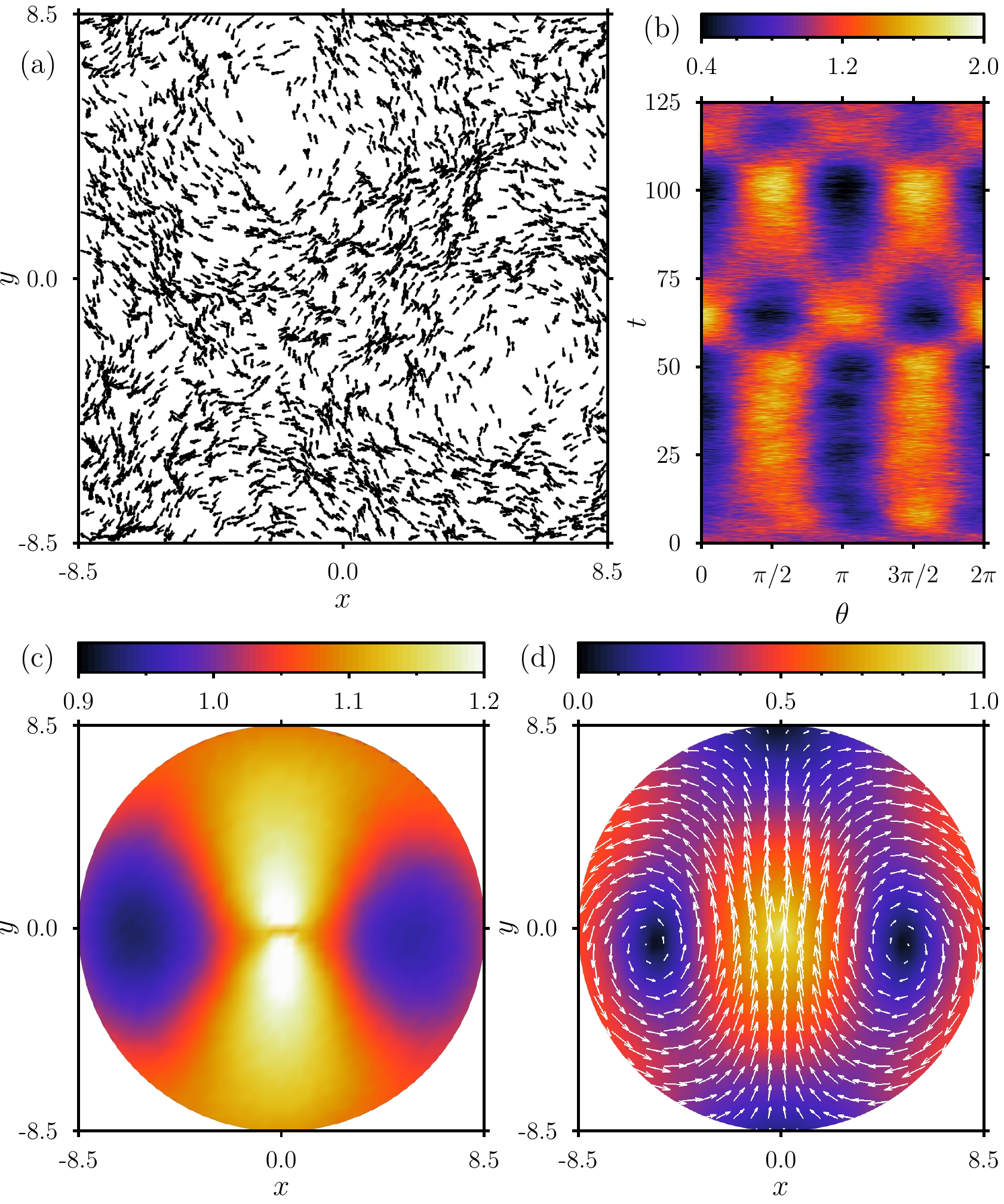}}
  \caption{(color online). Direct simulation with $N=3600$, $L/\ell=17$, $\phi=10\%$, and $\mathrm{Pe}=3.7$.\ (a) Snapshot of two counter-rotating vortices. (b) Temporal diagram of the mean orientation distribution, where $\theta=\cos^{-1}(\mathbf{p}\cdot\hat{\mathbf{x}})$.  (c)~Pair distribution function $g(\mathbf{r})$. (d) Pair polarization $\mathbf{P}(\mathbf{r})$. \label{fig:largetail}} \vspace{-0.2cm}
\end{figure}

We now turn to suspensions of large-tail swimmers ($\nu>0$), which, except at very high densities, are linearly stable \cite{Brotto13}.\ However, our simulations still reveal unusual dynamics for sufficiently positive values of $\mathrm{Pe}$.\ As shown in Fig.\ \ref{fig:largetail}\hyperref[fig:largetail]{(a)} and online video \cite{epaps}, these suspensions indeed tend to develop active `lanes', which often organize around pairs of large-scale counter-rotating vortices.\ As evidenced by the mean orientation distribution in Fig.\ \ref{fig:largetail}\hyperref[fig:largetail]{(b)} exhibiting two peaks separated by $\pi$, a strong nematic alignment, predominantly in the form of bend modes, exists at all scales including the system size.\ In addition, the temporal evolution of the orientations suggests the presence of persistent stationary structures that quasi-periodically form and break up over long periods. The pair distribution function in Fig.\ \ref{fig:largetail}\hyperref[fig:largetail]{(c)} shows a characteristic stripe in the longitudinal direction consistent with the presence of lanes, and the pair polarization still has a dipolar symmetry, with $\mathbf{P}$ now expectedly pointing in the opposite direction to that previously observed for large-head swimmers. 
\begin{figure}[t]
 \centerline{\vspace{-0.3cm}
  \includegraphics[scale=0.5]{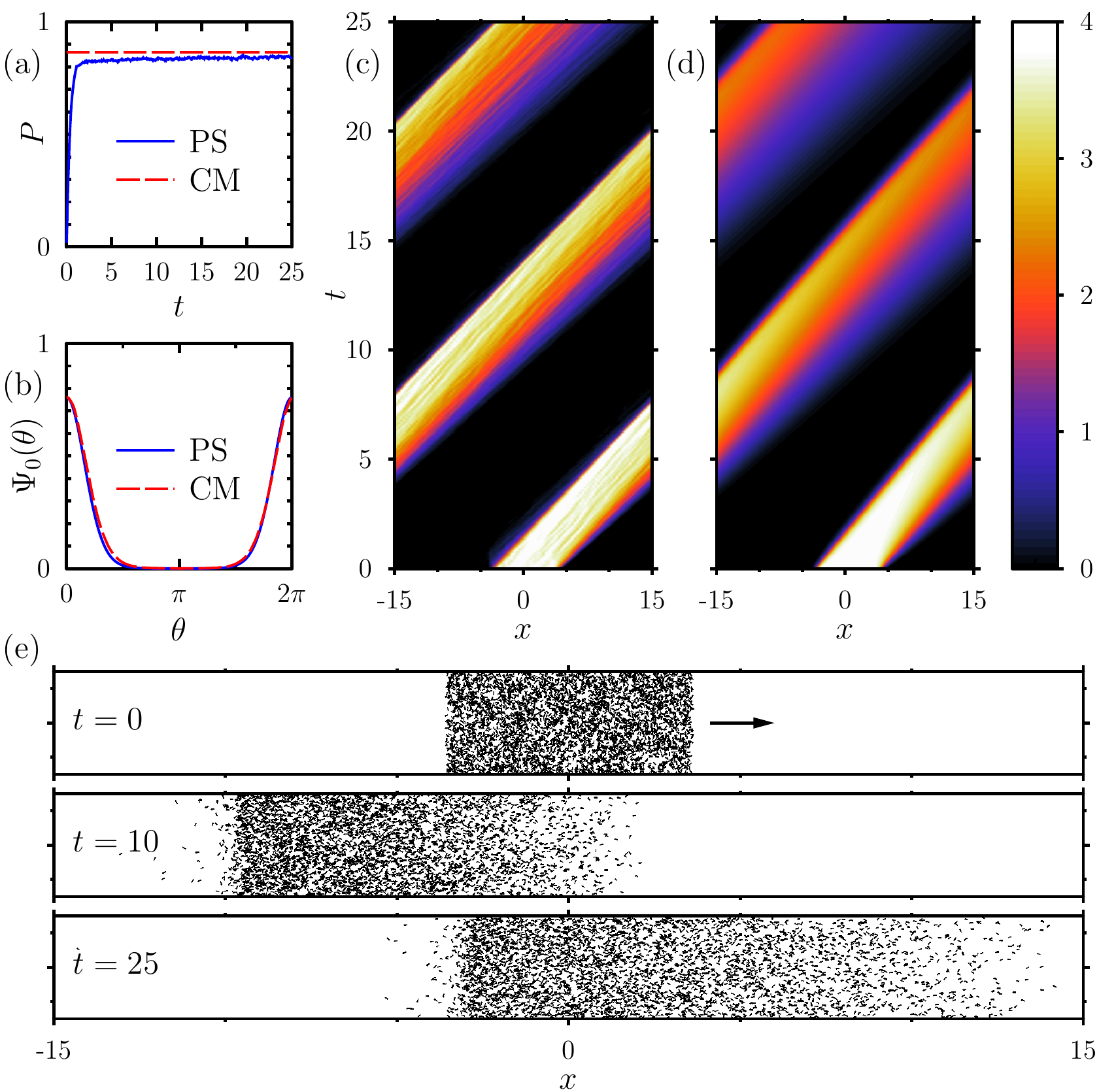}}
  \caption{(color online). Density waves in a narrow channel with a uniform flow. Comparison between particle simulation (PS) ($N=4000$, $\phi=0.05$, $\xi=4$, $L\times W=30\ell\times 3\ell$) and numerical solution of the continuum model \eqref{waveeq} (CM). (a) Mean polarization $P$ rapidly reaching $P_{0}$ (polar stable state). (b) Steady orientation distribution. (c)-(d) Spatiotemporal diagrams of longitudinal concentration from PS and CM, respectively. (e) Particle distributions (particles are magnified). \label{fig:waves}} \vspace{-0.4cm}
\end{figure}

Finally, we analyze the quasi-1D dynamics of a suspension confined in a narrow channel when a uniform external flow $\mathbf{U}_{0}=U_{0}\hat{\mathbf{x}}$ is applied in the longitudinal direction; see Fig.\ \ref{fig:waves}.\ A sufficiently strong flow can be shown to suppress the collective motions observed above by controlling particle alignment \cite{Lefauve13}, and orientations then approximately decouple from spatial concentration fluctuations: $\Psi(\mathbf{r},\mathbf{p},t)\approx c(x,t)\Psi_{0}(\theta,t)$, with $\theta=\cos^{-1}(\mathbf{p}\cdot\hat{\mathbf{x}})$. The steady anisotropic orientation distribution is easily obtained as $\Psi_{0}(\theta)=C\exp(\xi\cos\theta)$, where $C$ is a normalization constant and $\xi=\nu U_{0}/d$ a dimensionless flow strength, with a net longitudinal polarization $P_{0}=\int\Psi_{0}(\theta)\,d\theta>0$ as indeed observed in simulations [Figs.\ \ref{fig:waves}\hyperref[fig:waves]{(a)}-\hyperref[fig:waves]{(b)}].\ Assuming the form of interactions is unchanged by the boundaries, the Smoluchowski equation \eqref{smoluchowski} then simplifies to a 1D transport equation for $c(x,t)$:
\begin{equation}
\partial_{t}c+\partial_{x}[U_{0}c+v_{s}P_{0}(1-\sigma c)c]=D\partial^{2}_{xx}c. \label{waveeq}
\end{equation}
This quasilinear wave equation is commonly used as a basic model for traffic flow behavior \cite{Nagatani02}, where the coefficient $v_{s}P_{0}(1-\sigma c)$ can be interpreted as an Eulerian concentration-dependent velocity. The mean single-swimmer $x$-velocity $v_{s}P_{0}$ is renormalized by interactions: it is largest where $c(x)=0$ and decays linearly with $c$ to reach zero where $c(x)=1/\sigma$, the maximum concentration. Wave solutions of \eqref{waveeq} are characterized by the emergence of a shock at the rear and a rarefaction wave at the front, much like in a typical traffic jam. A shock and rarefaction wave are indeed observed in particle simulations in Figs.\ \ref{fig:waves}\hyperref[fig:waves]{(c)} and \hyperref[fig:waves]{(e)} and online video \cite{epaps}, showing the evolution of an initially uniform and isotropic finite-sized plug, and excellent agreement is found with a numerical solution of the traffic flow equation \eqref{waveeq} in Fig.\ \ref{fig:waves}\hyperref[fig:waves]{(d)}. Density shocks have previously been reported in experiments on driven confined emulsions, where the shocks occurred at the front of the waves and were a consequence of interactions with the lateral boundaries \cite{Beatus09,Desreumaux13}. While boundary interactions, which are taken into account in our simulations, still likely play a role in the present system, the primary mechanism here is the truly novel combined effect of active self-propulsion and renormalization by interparticle dipolar interactions.

We thank Denis Bartolo and Saverio Spagnolie for useful conversations, and gratefully acknowledge funding from NSF CAREER Grant CBET-1150590.

\end{document}